\documentstyle[sprocl,psfig]{article}

\def\be{\begin{equation}}
\def\ee{\end{equation}}
\def\bea{\begin{eqnarray}}
\def\eea{\end{eqnarray}}

\begin{document}

\hfill LBNL-47144
\vskip .01 in
\hfill December, 2000
\vskip .01 in

\title{Nonlinear QED Effects in Heavy Ion Collisions}
\author{Spencer R. Klein}

\address{Nuclear Science Division, Lawrence Berkeley
National Laboratory\\ Berkeley, CA, 94720, USA\\E-mail: SRKLEIN@LBL.GOV}


\maketitle

\abstracts{Peripheral collisions of relativistic heavy ions uniquely
probe many aspects of QED.  Examples include $e^+e^-$ pair production
and nuclear excitation in strong fields. After discussing these
reactions, I will draw parallels between $\gamma\rightarrow e^+e^-$
and $\gamma\rightarrow q\overline q$ and consider partly hadronic
reactions.  The scattered $q\overline q$ pairs are a prolific source
of vector mesons, which demonstrate many quantum effects.  The two
ions are a two-source interferometer, demonstrating interference
between meson waves. Multiple vector meson production will demonstrate
superradiance, a first step toward a vector meson laser.  Finally, I
will discuss the experimental program planned at the RHIC and LHC
heavy ion colliders.}

\centerline{Invited talk, presented at the}
\centerline{18th Advanced ICFA Beam Dynamics Workshop}
\centerline{ on Quantum Aspects of Beam Physics,} 
\centerline{October 15-20, 2000,  Capri, Italy}

\section{Introduction}

Heavy ion collisions might seem like a strange topic for an
accelerator physics conference.  However, many topics of interest to
accelerator physicists also occur in peripheral heavy ion collisions.
In these collisions, the ions do not physically collide.  Instead,
they interact electromagnetically at long ranges, up to hundreds of
fermi.  Relativistic heavy ions carry extremely strong electromagnetic
fields, allowing tests of nonperturbative electrodynamics.  These
fields are strong enough to allow for multiple reactions involving a
single pair of ions, so quantum fluctuations and superluminous
emission can be studied.  Even for single particle production, quantum
interference affects the vector meson spectrum.  All of these topics
have parallels in advanced accelerator design. And, some aspects of
heavy ion interactions impact directly on accelerator design.  This
writeup will review the physics of peripheral heavy ion collisions,
with an emphasis on principles.  Mathematical and experimental details
are left to the references.

\begin{figure}[t]
\centerline{\psfig{figure=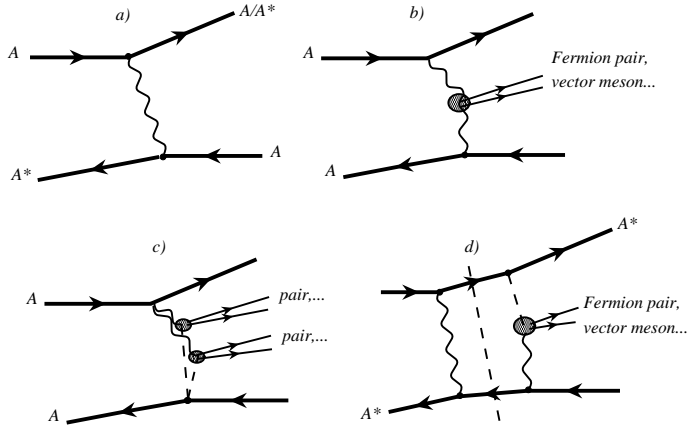,height=2.5 in}}
\caption[]{Some peripheral reactions: (a) Mutual nuclear
excitation. (b) Two-photon interactions (c) Multiple (double)
interaction, possible because $Z\alpha$ is large. (d) Two-photon
interaction with nuclear excitation.  The dashed line shows how the
reaction factorizes into independent two-photon (or photon-Pomeron)
and nuclear excitation reactions.  This is the dominant diagram; the
amplitude for excitation by the photon in (b) is
small~\cite{hencken}.}
\label{feynman}
\end{figure}

Several different types of peripheral reactions are possible.  The two
nuclei may exchange one or more photons (Fig \ref{feynman}a).  One or
both nuclei may be excited into a Giant Dipole Resonance (GDR) or
higher state. Or, the photon may interact with a single nucleon in the
nucleus in an incoherent photonuclear interaction.

Two fields may interact with each other.  In a two-photon interaction,
each nucleus emits a photon. The two photons collide to produce a
leptonic or hadronic final state, as in Fig. \ref{feynman}b.  The
fields are so strong that `two-photon' is a misnomer- the number of
photons from one nucleus may be large, and, in fact, poorly defined.
A photon from one nucleus may interact with the coherent meson or
Pomeron fields of the other.  Although this reaction has some
similarities with incoherent photonuclear interactions, coherence
restricts the final state kinematics, so reactions involving two
coherent fields produce kinematically similar final states.

Here, we (by definition) require that the two nuclei physically miss
each other and do not interact hadronically.  The impact parameter
$b>2R_A$, $R_A$ being the nuclear radius.  More detailed calculation
will calculate and use the non-interaction probability as a function
of $b$.

In the nuclear rest frame, a photon, Pomeron or meson coupling
coherently to a nucleus must have $p < \hbar c/R_A$. More precisely,
the coupling is governed by the nuclear form factor.  In a collider
where each nucleus is Lorentz boosted by $\gamma$, this coupling
transforms to $p_\perp < \hbar c/R_A$ and photon energy $k = p_{||} <
\gamma\hbar c/R_A$.  So, two-field interactions can occur up to a
maximum energy $W=2\gamma\hbar c/R_A$, with a final state $p_\perp <
2\hbar c/R_A$.  For photons, $p_\perp$ is actually smaller, peaked at
$p_\perp < \hbar c/b$.

Two-photon, photon-Pomeron/meson and double-Pomeron/meson reactions
are all possible.  Double-Pomeron/meson interactions are limited to a
narrow range of impact parameter because of the short range of the
strong force.  Therefore, they will occur with a relatively low cross
section.  They will also have a quite different $p_\perp$
spectrum. The $p_\perp$ spectral difference will allow some
statistical separation between two-photon and photon-Pomeron
interactions.

For most applications, the electromagnetic fields of
ultra-relativistic nuclei may be treated as a field of virtual
photons, following Weizs\"acker-Williams.  The photon flux 
from a nucleus with charge $Z$ a distance
$r$ from a nucleus is
\begin{equation}
{d^3N(k,r) \over dkd^2r} = {Z^2\alpha x^2 \over \pi^2kr^2} K_1^2(x)
\end{equation}
where $x=kr/\gamma\hbar$ and $K_1(x)$ is a modified Bessel
function. The two-photon luminosity is the overlap of the two photon
fields. The usable two-photon luminosity $L_{\gamma\gamma}$ is this
overlap, integrated over all $b>2R_A$.  This can be calculated
using
\begin{equation}
\begin{array}{rcl}
L_{\gamma\gamma} (W,Y)&  = & L_{AA} \int{dk_1\over k_1} \int {dk_2\over k_2} \\
& &  2\pi \int_{R_A}^\infty b_1db_1 
\int_{R_A}^\infty b_2db_2 \int_0^{2\pi} d\phi 
{d^3N(k_1,b_1)\over dk_1d^2b_1}
{d^3N(k_2,b_2)\over dk_2d^2b_2}
\Theta(b-R_1-R_2)
\end{array}
\label{eq:gglum}
\end{equation}
where $L_{AA}$ is the nuclear luminosity, $\Theta$ is the step
function and the impact parameter
$b=\sqrt{(b_1^2+b_2^2-2b_1b_2cos(\phi))}$~\cite{gglumbaur}
\cite{gglumcahn}.  This must be evaluated numerically.  The
requirement that the nuclei not physically collide ($\Theta$ function)
reduces the flux by about 50\%.  The final state energy $W=4k_1k_2$
and rapidity $y=1/2\ln(k_1/k_2)$ can also be found.  Usually, the
slight photon virtuality $q^2 < (\hbar/R_A)^2$ can be neglected.  The
exception is $e^+e^-$ production, since $q^2 \sim (\hbar/R_A)^2 \gg
m_e^2$.

Since Pomerons and mesons are short-ranged, photon-Pomeron and
photon/meson interactions take place inside one of the nuclei.  At a
given $b$, the photon intensity is found by integrating the photon
flux over the surface of the target nucleus, and normalizing by
dividing by the area $\pi R_A^2$.  The total effective photon flux is
this intensity, integrated over all $b>2R$.  It is found analytically;
the result is within 15\% of the integrated flux in the region
$b>2R_A$:
\begin{equation}
{dN_\gamma \over dk} = {2Z^2\alpha\over\pi k} \big(XK_0(X)K_1(X) - 
{X^2\over 2} [K_1^2(X) - K_0^2(X)\big)
\end{equation}
where $X=2R_A k/\gamma$.  For $X<1$, the total number of photons
with $k_{min} < k < k_{max}$ is
\begin{equation}
N_\gamma =  {2Z^2\alpha\over\pi } \ln\big({k_{max} \over k_{min}}\big).
\end{equation}
For photo-nuclear interactions, the maximum photon energy seen by one
nucleus is strongly boosted, by $\Gamma=2\gamma^2-1$, or 20,000 for
RHIC and $1.5\times10^7$ for LHC.  Thus, the photon energies
reach 600 GeV with gold at RHIC, and 500 TeV for lead at the LHC;
with lighter nuclei, these numbers are 2-3 times higher.

\begin{table}
\caption{Beam Species, Energies, Luminosities, compared for
RHIC (Summer, 2000), RHIC Design and LHC.  RHIC is expected to
reach it's design parameters in 2001.}
\vspace{0.2 cm}
\begin{center}
\begin{tabular}{|l|r|r|r|}
\hline
Machine       & Species    & Beam Energy & Max. Luminosity \\
               &            & (per nucleon) & (cm$^{-2}$s$^{-1}$) \\
\hline

RHIC 2000 & Gold & 65 GeV   & $2\times 10^{25}$ \\
RHIC & Gold    & 100 GeV    & $2\times 10^{26}$ \\ 
RHIC & Silicon & 125 GeV    & $4.4\times 10^{28}$ \\ 
LHC  & Lead    & 2.76  TeV  & $1\times 10^{26}$ \\ 
LHC  & Calcium & 3.5  TeV   & $2\times 10^{30}$ \\
\hline
\end{tabular}
\end{center}
\end{table}

Fixed target heavy ion accelerators can produce $e^+e^-$ pairs, with
and without capture; heavier states are not energetically accessible.
These reactions have been studied at the LBL Bevalac, BNL AGS and CERN
SPS.  Studies of hadroproduction is just beginning at the Relativistic
Heavy Ion Collider (RHIC) at Brookhaven National Laboratory, and the
Large Hadron Collider at CERN; these colliders are energetic enough to
produce a variety of final states.  The characteristics of these
colliders are shown in Table 1.

Peripheral collisions have recently been reviewed
by Baur, Hencken and Trautmann~\cite{baurrev}.

\section{Nuclear Excitation and Incoherent Photonuclear Interactions}

For low energy photons, nuclear excitations are typically collective.
For example, in a Giant Dipole Resonance, the protons oscillate in one
direction and the neutrons in the other.  This vector oscillation can
be induced by a single photon.  Higher excitations include double (or
higher) Giant Dipole Resonances, higher $n$ states of a harmonic
oscillator.  There are also Giant Quadrupole Resonances, which require
multiple photons to produce.  These states typically decay by emitting
one or more neutrons which can be detected in far-forward
calorimeters.

These reactions are of interest for a couple of reasons.  As Table 2
shows, the cross sections are substantial~\cite{dissoc}.  Nuclear
excitation is a substantial contributor to beam loss.  The photon
carries little momentum, so nuclear excitation creates a beam of
particles with unchanged momentum but altered charge to mass
ratio~\cite{beampipe}.  This beam will escape the magnetic optics and
strike the beampipe at a relatively well defined point downstream,
locally heating the magnets. This heating could cause superconducting
magnets to quench.   Also, this beam could be extracted from the
accelerator, for fixed target use.  

A single photon can excite both the emitting and target nuclei,
although the cross section is smaller than for single excitation. This
double process is significant for a couple of reasons.  It has a clean
signature and is useful as a luminosity monitor~\cite{lummeas}.
Second, it can tag small $b$ events.  To a good approximation, the
nuclear excitation photon factorizes from the remainder of the
interaction~\cite{hencken}, as is shown in Fig. 1(d).  Thus the
nuclear excitation can tag collisions at low $b$.

\begin{table}
\caption[]{Cross sections for nuclear excitation~\cite{beampipe}, pair
production (Eq. \ref{eq:pert}), bound-free pair
production\cite{beampipe}, $\rho$, $J/\psi$
and double-$\rho$ production~\cite{vmprod}.  The nuclear excitation and
bound $e^-$ cross sections are per ion.}
\vspace{0.2 cm}
\begin{tabular}{|l|r|r|r|r|r|r|}
\hline
System & $\sigma(Exc.)$ & $\sigma(e^+e^-)$ & 
$\sigma($bound $e^-)$ & $\sigma(\rho)$ &
$\sigma(J/\psi)$ & $\sigma(\rho\rho)$ \\
\hline
RHIC-Au & 58 b   & 33 kb & 45 b   & 590 mb & 290 $\mu$b & 720$ \mu$b\\
RHIC-Si & 150 mb & 41 b & 1.8 mb & 8.4 mb & 3.6 $\mu$b &  \\
LHC-Pb  & 113 b  & 150 kb & 102 b  & 5.2 b & 32 mb  & 8.8 mb\\
LHC-Ca  & 800 mb & 600 b & 36 mb  & 120 mb & 390 $\mu$b &  \\
\hline
\end{tabular}
\end{table}

\section{Two-Photon Interactions}

Two-photon interactions have been studied extensively at $e^+e^-$
colliders.  Photons couple to charge, so two-photons couplings measure
the internal charge content of mesons; $q\overline q$ pairs are
produced, but not charge-free states like glueballs.  Hybrids
($q\overline qg$) and 4-quark states ($q\overline qq\overline q$) are
produced at intermediate rates.  Thus, coupling to two-photons is a
sensitive test for exotic mesons.

Meson pair production rates depend on the pair energy.  Near
threshold, charged meson pairs ($\pi^+\pi^-$) are produced, but
neutral pairs ($\pi^0\pi^0$) are not.  At higher energies, the photons
see the quark structure of mesons, and both charged and neutral mesons
are produced.

\begin{figure}[t]
\centerline{\psfig{figure=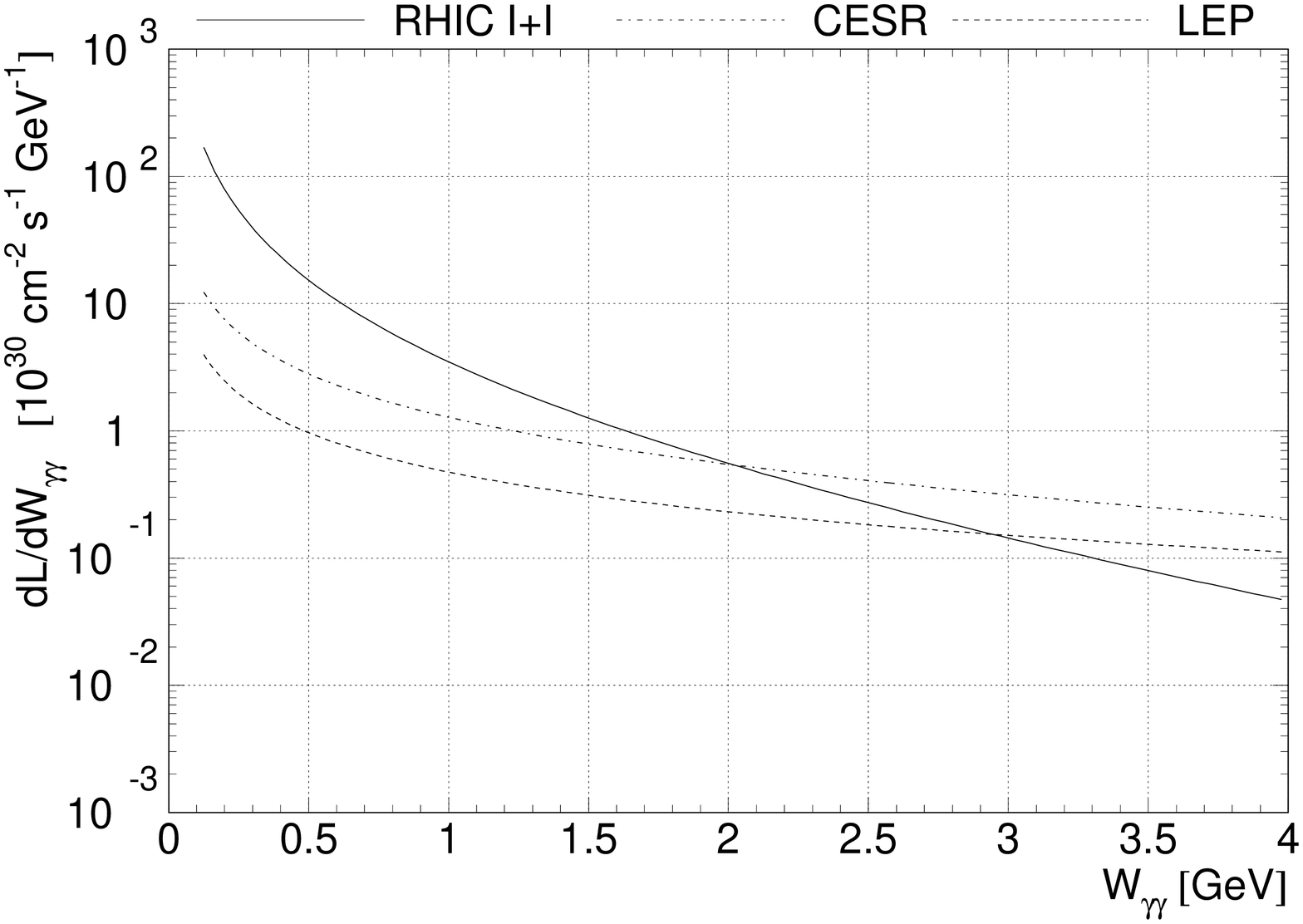,height=2 in}}
\caption{Two-photon luminosity expected at RHIC with gold and iodine
beams, compared with the luminosities at LEP II ($\sqrt{s}=$ 180 GeV
and a luminosity of $5\times10^{31}$cm$^{-2}$s$^{-1}$) and at CESR
($\sqrt{s}=$ 10 GeV and a luminosity of
$2.9\times10^{32}$cm$^{-2}$s$^{-1}$).}
\label{luminosity}
\end{figure}

Two-photon interactions at heavy ion colliders are of interest because
that the luminosity scales as $Z^4$ and extremely high rates are
possible. Figure \ref{luminosity} compares the $\gamma\gamma$
luminosities at RHIC, with the LEP and CESR $e^+e^-$
colliders~\cite{lund}; for $W<1.5$ GeV, RHIC can reach the highest
presently available two-photon luminosities. Heavy ion colliders also
probe some unique areas, such as multiple pair production, and
bound-free pair production; both are probes of strong field QED.

\subsection{Lepton Pair Production}

Lepton pair production can test the limits of perturbative QED.
Perturbation theory may fail because the coupling $A\alpha$ is so
large.  Even with perturbative approaches, $e^+e^-$ production
introduces additional complications.  The electron Compton wavelength,
$\Lambda_e$=386 fm, is large compared to typical impact parameters. So
at $W\sim 2 m_e$, where the bulk of the cross section is, the pair
production is poorly localized.

The first perturbative calculation specific to heavy ion collisions was
by Bottcher and Strayer~\cite{bottcher}.  They treated the ions as
sources of classical (but relativistic) electromagnetic potentials
that follow fixed trajectories.  This approach naturally incorporated
off-shell photons.  This calculation also accounted for large electron
Compton radius $\Lambda_e = 386$ fermi, with an appropriate cutoff.
In the two-photon approach, $\Lambda_e$ should replace the minimum
impact parameter, $R_A\sim 7$ fermi, in Eq. \ref{eq:gglum}.  This
reduces the cross section significantly compared to earlier
calculations.

A slightly later, more refined calculation by Baur and Bertulani
included Coulomb corrections, to account for the fact that the pair is
produced deep in a Coulomb potential~\cite{firstrev}.  With this
refinement, the cross section is given by
\begin{equation}
\sigma = {28Z^4\alpha^4 \hbar^2 \over 27\pi m_e^2 c^2} 
\big(\ln^3({\Gamma\delta\over 2}) - {3\over 2} (1+ 2\overline f)
\ln^2({\Gamma\delta\over 2})\big)
\label{eq:pert}
\end{equation} 
where $m_e$ is the electron mass, $\delta=0.681$ is Euler's constant
and $\overline f = (Z\alpha)^2 $ $\Sigma_{n=1}^\infty
[n(n^2+Z^2\alpha^2)]^{-1}$ is the usual Coulomb correction.  The
$\ln^3$ term dominates at high energy.  Other authors have found
slightly different results, depending on the details of the
calculation.

Baur and Bertulani also calculated the probability of pair production
at a given $b$.  With gold at RHIC, this probability is greater than 1
for $b=\Lambda_e$!  The differential cross section $d\sigma/2\pi bdb$
saturates.  The problem is resolved by multiple pair production: a
single ion pair small-$b$ confrontation can produce more than one
pair.  The the number of pairs is Poisson distributed, with the
$b$-dependent mean~\cite{multipair}.  This saturation can also affect
calculations of the single pair cross section.

Numerous authors have considered non-perturbative $e^+e^-$ production,
usually using the time-dependent Dirac equation.  Some authors solved
the coupled-channel equations numerically.  The ions were stepped
through their positions. At each step, the coupling from the initial
state to a pair-containing final states was calculated.  An accurate
calculation requires a complete and orthogonal set of states.  This
turned out to be rather difficult, and early calculations found
results that varied by orders of magnitude.

Baltz and McLerran calculated pair production to all
orders~\cite{Baltz}.  Their method is similar to the perturbative
calculation.  They worked in light-cone coordinates with
Lienard-Wiechert potentials similar to those of Bottcher \&
Streyer. They first found the Greens function for the exact wave
function at the interaction point.  The transition amplitude was then
constructed from the Greens function. The total cross section is this
amplitude, integrated over impact parameter and intermediate
transverse momentum.  Their result matches the perturbative result
(without Coulomb corrections).

Recently, Roman Lee and A. Milstein found a problem with the order of
integration in the Baltz and McLerran paper~\cite{lee}.  When the order
changed, Lee and Milstein the result changed to include the Coulomb
correction found by Baur \& Bertulani (the $\overline f$ term in
Eq. \ref{eq:pert}).

The agreement with perturbation theory is somewhat surprising, given
the large coupling. However, Baltz and McLerran found that, for
multiple pair production, their result was smaller than the
perturbative result.  Since multiple pair production is naturally a
higher order process, it's not surprising that a difference appears.

A related reaction is bound-free pair production where the electron is
produced bound to one of the nuclei.  As with free pairs, perturbative
calculations may be inadequate, and an exact solution to the
time-dependent Dirac equation is desired.  This problem has also been
tackled perturbatively; here the final state consists of a free
positron and an electron in an atomic orbital.  The cross section to
produce an electron bound in an atomic $K-$ shell is~\cite{baurrev}
\begin{equation}
\sigma = {33\pi Z^8\alpha^8\hbar^2 \over 10 m_e^2c^2}
{1 \over exp(2\pi Z\alpha) -1 }
\big[\ln({\Gamma\delta\over 2}) - {5\over 3}\big].
\label{eq:capture}
\end{equation}
The stronger $Z$ dependence comes from the electron-nucleus binding
energy.  Inclusion of higher shells will increase this by about 20\%.
This cross section has the form $\sigma = A\ln(\gamma) + B$.
Extrapolations from lower energy data using this form find a cross
section about twice as large~\cite{grafstrom}.  Coupled-channel
calculations have been tried on this problem, and produced a wide
range of results.  Also, as with free-production, an all-order
solution to the time-dependent Dirac equation has recently been found,
again using light-cone coordinates~\cite{baltz2}.  The result was
slightly lower than perturbation theory.  The cross section for
bound-free production is much lower than for free production,
so that $d\sigma/2\pi bdb$ is not saturated. 

The 1-electron atoms produced in this reaction have their momentum
unchanged, so that they will follow well-defined trajectories.
As with nuclear excitation, this can lead to heating of the
accelerator magnets and also allow for extracted beams~\cite{beampipe}.

In principle, these non-perturbative aspects of pair production also
apply to $\mu^+\mu^-$ and $\tau^+\tau^-$ production.  However, the
masses are much larger, so any non-perturbative effects are much
smaller.  Because $m_\mu > \hbar/R_A$, Eq. \ref{eq:gglum} applies for
heavy lepton production.

\section{$q\overline q$ fluctuations and Vector Meson Production}

The vacuum fluctuation $\gamma\rightarrow q\overline q$ is similar to
$\gamma\rightarrow e^+e^-$; only the final state charges and masses
are different.  Just as the virtual $e^+e^-$ pair can interact with an
external Coulomb field and become real, the $q\overline q$ pair can
interact with an external nuclear field and emerge as real vector
meson~\cite{vmprod}.

This picture is clearest in the target rest frame.  The incoming
photon has a high momentum, and the fluctuation persists for a time
$\tau_f=\hbar/M$, during which it travels a distance known as the
formation distance $l_f = 2\hbar k/M^2$.  In alternate language,
$l_f=\hbar/p_{||}$, where $p_{||}$ is the momentum transfer required
to make the pair real. For $e^+e^-$ pairs, $l_f$ is typically much
larger than a single atom; for $q\overline q$ pairs, $l_f$ is
typically much larger than a single nucleus.  So, the fluctuation
cannot see the target structure. During it's lifetime, the fluctuation
can interact with the external field to become a real pair.

The $q\overline q$ scatters elastically from the a nucleus with atomic
number $A$. This scattering is mediated by the strong force and
transfers enough momentum to give the meson its mass. The scattering
leaves the photon quantum numbers $J^{PC}$ unchanged.  This elastic
scattering cannot easily be described in terms of quarks and gluons.
The most successful description is in terms of the
Pomeron~\cite{forshaw}.  For hard processes the Pomeron may be thought
of as a 2-gluon (quasi-bound) ladder, connected by gluon rungs.
However, for soft processes such as elastic scattering, this picture
may be inappropriate.  For soft reactions, the best picture is the
40-year old soft-Pomeron diffractive picture~\cite{review}.  The
Pomeron absorbs part of the photon wave function, allowing a
$q\overline q$ to emerge dominant.

In this model, the cross section for the reaction $ A + A \rightarrow
A + A + V$ may be calculated in a straightforward manner.  The
starting point is data on $\gamma + p \rightarrow V + p$ from fixed
target experiments and HERA.  The forward scattering amplitudes may be
parameterized $d\sigma/dt|_{t=0} = b_v(XW^\epsilon + YW^{-\eta})$,
where $t$ is the 4-momentum transfer from the nucleus and here $W$ is
the $\gamma p$ center of mass energy.  The first term, with $\epsilon
\sim 0.22$, is for Pomeron exchange, while the second is for meson
exchange; Pomeron exchange dominates at high energies.  This amplitude
factorizes into two parts: the $\gamma\rightarrow q\overline q$
amplitude and the elastic scattering amplitude.  The first part can be
determined from the partial width for $V\rightarrow e^+e^-$, allowing
vector meson production data to fix the scattering amplitude.  Vector
meson dominance allows us to treat the $q\overline q$ fluctuation as a
real vector meson.  The optical theorem can be used to find the total
$Vp$ cross section.

The total $VA$ cross section may be found with a Glauber calculation.
This calculation integrates over the transverse plane, summing the
probability of having 1 or more interactions:
\begin{equation}
\sigma_{tot}(VA) = \int d^2{\vec r} \big(1 -
e^{-\sigma_{tot}(Vp)T_{AA}({\vec r})} \big)
\end{equation}
where $T_{AA}({\vec r})$ is the nuclear thickness function.
These cross sections rise with $W$ at low energies, then level off
at an almost constant value.   

The optical theorem is used to find $d\sigma/dt|_{t=0}$ for the meson
-nucleus scattering.  Finally, the leptonic width is used to find the
forward amplitude for vector meson production.  In the small-$\sigma$
limit, $\sigma_{tot}(Vp)T_{AA}(b=0)\ll 1$, the forward amplitude
scales as $A^2$.  This limit applies for heavy systems such as
$c\overline c$.  As $\sigma_{tot}(Vp)$ rises, the $A-$dependence
decreases, and for large $\sigma_{tot}(Vp)$, the scaling is $A^{4/3}$,
with the vector meson seeing the front face of the nucleus.

The total photonuclear cross section is given by an integration over
$t$:
\begin{equation}
\sigma(\gamma A\rightarrow VA) = d\sigma/dt(\gamma A\rightarrow VA)|_{t=0}
\int_{t_{min}}^\infty dt |F(t)|^2
\label{eq:siggamma}
\end{equation}
where $t_{min}=M_v^2/4k$ and $F(t)$ is the nuclear form factor.  For a
heavy nucleus, $F(t)$ may be fit analytically by a convolution of a hard
sphere with a Yukawa potential.

Eq. \ref{eq:siggamma} agrees well with data from fixed target
experiments.  The total cross section is
\begin{equation}
\sigma(A + A \rightarrow A + A + V) = 2 \int dk {dN_\gamma \over dk}
\sigma(\gamma A\rightarrow VA). 
\label{eq:sigmatot}
\end{equation}
The factor of 2 is because either nuclei can act as target or emitter.
These cross sections are given in Table 2.

The implications of this straightforward calculation are significant.
The cross sections are huge.  With gold at RHIC, $\rho^0$ production
is 10\% of the total hadronic cross section.  With lead at LHC, the
$\rho^0$ cross section is about equal to the hadronic cross section!
Heavy ion colliders can act as vector meson factories, with rates
comparable to $e^+e^-$ vector meson machines.  The $10^{10}$ $\phi$
produced in $10^6$ seconds with calcium beams at LHC is comparable to
that expected at a dedicated $\phi$ factory.  Searches for rare decay
modes, CP violation and the like are possible.  Also, vector meson
spectroscopy will be productive; mesons like the $\rho(1450)$,
$\rho(1700)$ and $\phi(1680)$ will be copiously produced.

\begin{figure}[t]
\centerline{\psfig{figure=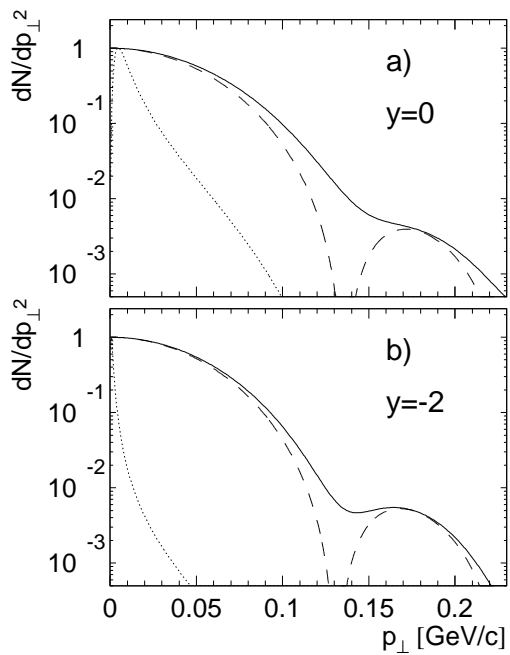,height=3.5 in}}
\caption{The vector meson $p_\perp$ spectrum (solid line) at
$y=0$ (a) and $y=2$ (b) is the convolution of the 
photon $p_\perp$ (dotted line) and the scattering
$p_\perp$ transfer (dashed line).}
\label{mesonpt}
\end{figure}

Fully coherent final states will be distinctive.  The final state
$p_\perp$ is a convolution of the photon and Pomeron $p_\perp$.
Figure \ref{mesonpt} shows these contributions.  The mean $p_\perp$
from the photon is $\hbar/b$, considerably smaller than $\hbar/R_A$.

\begin{figure}[t]
\centerline{\psfig{figure=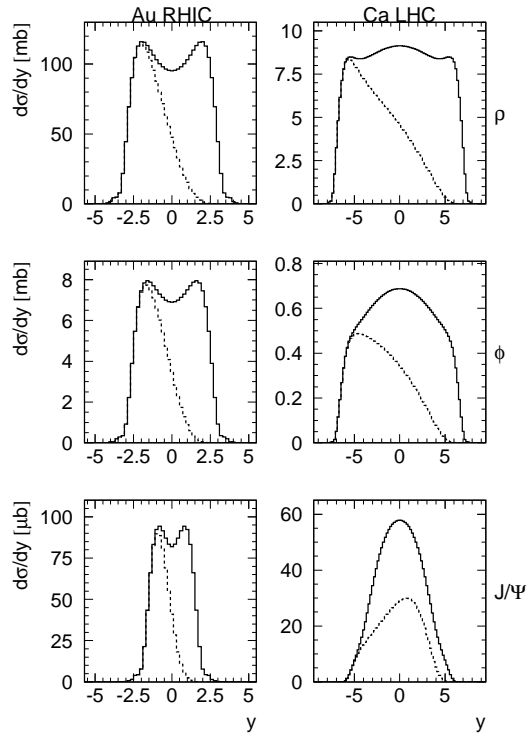,height=4 in}}
\caption{Rapidity distribution $d\sigma/dy$ with gold
at RHIC (left panels) and calcium at the LHC (right panels)
for the $\rho^0$, $\phi$ and $J/\psi$.  The solid line is the
total, while the dashed line shows the production for a single
photon direction.}
\label{mesony}
\end{figure}

This approach can also find the vector meson rapidity distribution.
The final state rapidity $y=1/2 \ln(M_V/k)$.  So, $d\sigma/dy = k/2
d\sigma/dk$ and can be determined from Eq. \ref{eq:sigmatot}.  The
photon can come from either direction, so the total $\sigma(y)$
includes contributions for photons from $+y$ and $-y$.  $d\sigma/dy$
is shown in Fig.~\ref{mesony}.

\subsection{Interference}

The observed $p_\perp$ spectrum is more complicated than Fig. 3
shows. Either nucleus can emit the photon.  The two possibilities are
indistinguishable, and therefore, they interfere.  In essence, the two
nuclei act as a two-source interferometer.  The two possible emitters
are related by a parity transformation.  Vector mesons are negative
parity so the two possibilities contribute with opposite signs,
producing destructive interference~\cite{interfere}. 
The cross section is
\begin{equation}
\begin{array}{rcl}
\sigma(p_\perp,y,b) &  = & A^2(p_\perp,y,b) +   A^2(p_\perp,-y,b) \\
& - & 2 A(p_\perp,y,b)A(p_\perp,-y,b) \cos(\phi(y)-\phi(-y) + 
{\vec p_\perp}\cdot {\vec b})
\end{array}
\end{equation}
where $A(p_\perp,-y,b)$ is the production amplitude and $\phi(y)$ is
the production phase.  $A$ may be found from the previous section.
For pure Pomeron exchange, the production is almost real.  The
production phase always cancels at $y=0$, and cancels everywhere
unless $\phi$ depends on $k$.  Variation is likely with the $\rho$ and
$\omega$ because of the meson contribution.  For other mesons, it is
likely to be small or negligible.

At midrapidity, the interference simplifies to 
\begin{equation}
\sigma(p_\perp,y=0,b) = A^2(p_\perp,y=0,b)(1-cos[\vec{p}\cdot\vec{b}]).
\end{equation}
For a given $b$, $\sigma$ oscillates with period $\Delta
p_\perp=\hbar/b$.  When $p_\perp b < \hbar$, the interference is
destructive and there is little emission.  The mean $b$ for $\rho$
production at RHIC is about 40 fermi, rising to 300 fermi at LHC.

\begin{figure}[t]
\centerline{\psfig{figure=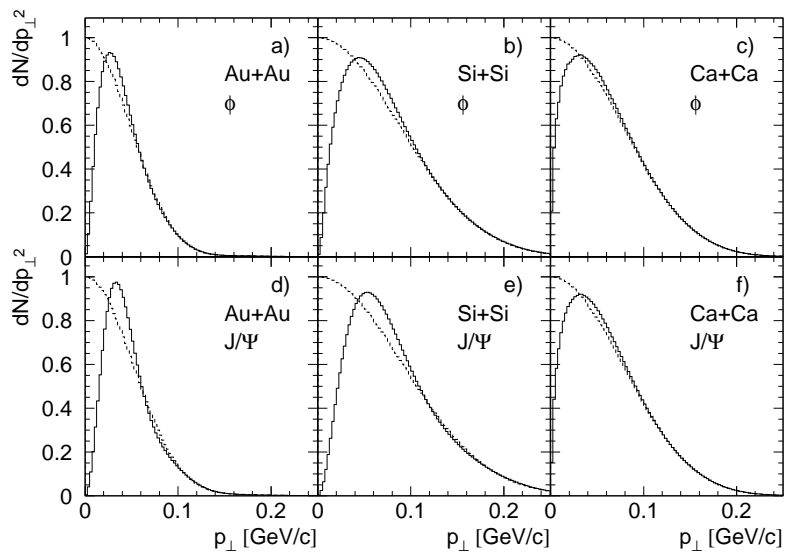,height=3 in}}
\caption{Meson $p_\perp$ spectra, with (solid lines) and without
(dashed line) interference, at y=0. The top panels are for the $\phi$,
and the bottom for the $J/\psi$, with gold (left) and silicon (center)
at RHIC, and calcium at the LHC (right).}
\label{mesonint}
\end{figure}

The impact parameter is unmeasured, so it is necessary to integrate
over all $b$.  This dilutes the interference, except for
$p_\perp<\hbar/\langle b\rangle$.  Figure \ref{mesonint} shows the
expected $p_\perp$ spectrum with and without interference.

The mean impact parameter for $\rho$ production with gold at RHIC is
40 fermi, far larger than the rho decay distance $\gamma\beta c\tau <
1$ fermi.  The vector mesons decay before their wave functions can
overlap! However, the decay product do overlap and interfere.  The
angular distributions for the two $\rho^0$ sources are the same, so
the interference pattern is not affected.  This process requires a
non-local wave function.  Consider $\rho^0\rightarrow\pi^+\pi^-$, with
$b\sim 40$ fermi.  Before the $\pi^+$ waves from the 2 sources can
overlap, they must travel $\sim$ 20 fermi each, during which time the
$\pi^-$ waves will travel 20 fermi in the opposite direction, and the
$\pi^+$ and $\pi^-$ waves will be separated by 40 fermi.  So,
non-locality is required to produce this interference pattern.

Although there is as yet no counterpart to Bell's inequality, the
choice of quantum observable does matter for this system.  Consider a
system where $b$ is measured.  For the $\pi^+$ and $\pi^-$, one can
measure either the momentum or position.  If the momenta of both $\pi$
are measured, then the interference pattern is observed.  If the
$\pi^+$ momentum is known, that disallows certain values of $\pi^-$
momentum where destructive interference is complete.  If the positions
of both $\pi$ are measured, the production point can be determined,
but the interference disappears.  If one position and one momentum are
observed, neither the interference pattern nor the production point
can be determined.

The wave function of the system is
\begin{equation}
\Psi(\vec{x}) = \exp(i(\vec{k}_-+\vec{k}_+)\cdot\vec{x})
\big[ \exp(i(\vec{k}_-+\vec{k}_+)\cdot\vec{R}_A) - 
 \exp(i(\vec{k}_-+\vec{k}_+)\cdot\vec{R}_B)\big]
\end{equation}
where $\vec{x}$ is where the vector meson would be if it didn't decay;
in the vector meson rest frame $\vec{x}= 1/2 (\vec{x}_+ + \vec{x}_-)$
where $\vec{x}_+$ and $\vec{x}_-$ are the position for the $\pi^+$ and
$\pi^-$, and $\vec{k}_+$ and $\vec{k}_+$ their momenta.  This wave
function cannot be factorized: $\Psi(\pi^+\pi^-) \ne
\Psi(\pi^+)\Psi(\pi^-)$.  Since the $\pi^+$ and $\pi^-$ are well
separated, the wave function is non-local.  This system is thus an
example of the Einstein-Podolsky-Rosen paradox.

\subsection{Multiple Vector Meson Production}

\begin{figure}[t]
\centerline{\psfig{figure=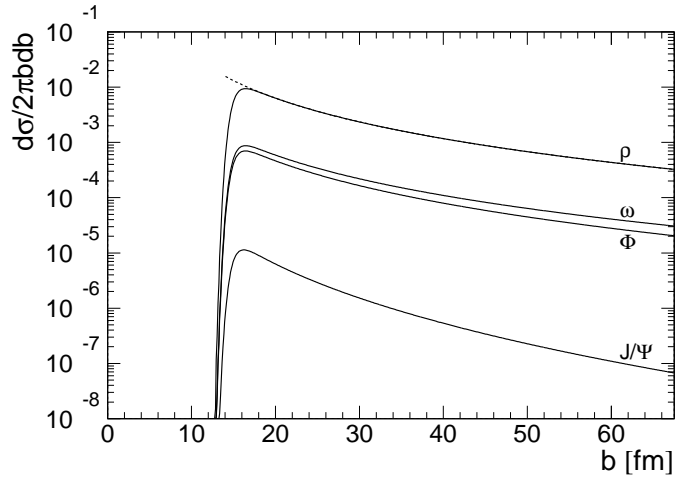,height=2.5 in}}
\caption{Probability of meson production, with gold at RHIC, as a
function of $b$.}
\label{mesonprob}
\end{figure}

The vector meson production probability at a given $b$ may be
calculated with the impact-parameter dependent photon flux.  This is
shown in Fig. \ref{mesonprob}.  At $b=2R$, the probability of $\rho^0$
production is 1\% at RHIC, rising to 3\% at LHC.  These probabilities
are high enough that multiple meson production should be observable.
In the absence of quantum or other correlations, multiple meson
production should be independent and Poisson distributed. At $b=2R$,
the $\rho^0\rho^0$ probabilities are $(1\%)^2/2$ and $(3\%)^2/2$ at
RHIC and LHC respectively. After integration over $b$, 1.4 million
$\rho^0\rho^0$ are expected per year at RHIC.  Like meson triples
should also be produced in observable numbers. Vector mesons are
bosons so production of like-meson pairs should be enhanced for
momentum differences $\delta p < \hbar/R_A$.  The meson follows the
photon spin and can be aligned or anti-aligned with the beam
direction, so the enhancement is only 50\%, so N(pair)$\sim = 1 + 0.5
\exp(\delta p R_A/\hbar)$.

\section{Experimental Status}

Fixed target measurements have been published for pair production, with
and without capture, and nuclear excitation.  Due to space
limitations, this writeup will only consider relativistic collisions,
with $\Gamma > 10$.  The solid targets, with the nuclei surrounded by
their electron clouds, differ from the stripped ion collisions we
focus on here.  Measurements of pair production in sulfur on
heavy ion collisions around $\Gamma=160$ have matched theoretical
predictions~\cite{pairprod}.  Pair production with capture has also
been studied with lead beams~\cite{grafstrom}.  As was previously
mentioned, when scaled to RHIC and LHC energies, this data may exceed
current estimates.  However, corrections may be needed for the limited
boost of the current experiments.

Programs to study a variety of peripheral reactions are underway in
the STAR collaboration at RHIC and the CMS collaboration at LHC.  For
most reactions, the largest backgrounds are expected to be grazing
hadronic collisions, beam gas interactions, and incoherent
photonuclear interactions~\cite{lund}.  For triggering, debris from
upstream interactions, and cosmic ray muons can be important.

These backgrounds can be separated from the signals by selecting
events with low multiplicity, typically, 2 or 4, low total $p_\perp$,
and zero net charge.  Baryon number and strangeness must also be
conserved. 

At the trigger level, significant rejection can be achieved by
requiring that the event originate inside the interaction region;
this removes most of the beam gas events, along with almost all
of the upstream interactions and cosmic ray muons.  Event timing
cuts also help reject cosmic ray muons.  

The STAR detector combines a large acceptance with a flexible
trigger~\cite{howard}.  Charged particles are detected in the
pseudorapidity range $|\eta|<2$ and $2.4 < |\eta| < 4$ by a large
central time projection chamber (TPC) and two forward TPCs.  This TPC
can also identify particles by $dE/dx$.  Neutral particles are
detected by a central barrel ($|\eta| < 1$) and endcap ($1 < \eta <
2$) calorimeter.  Two zero degree calorimeters will detect neutrons
from nuclear breakup, useful for background rejection.

For triggering, a scintillator barrel covering $|\eta|<1$ and
multi-wire proportional chambers covering $1< |\eta|< 2$ measure
charged particle multiplicity on an event by event basis.  These
detectors have good segmentation, allowing for total multiplicity and
topological selection in the trigger.  The trigger has 4 levels, with
the earliest level based on field programmable gate arrays and the
later levels computer based.  The final selection uses on-line TPC
tracking.  Peripheral collisions data will be collected in parallel
with central collision data.  Simulations show that the planned
trigger algorithms should be able to efficiently select peripheral
events while rejecting enough background enough to minimize
deadtime~\cite{lund}.

STAR took it's first data this summer (2000).  The central TPC,
scintillator barrel and zero degree calorimeters were operational.
Although the trigger was not completely functional, in late August,
the collaboration took about 7 hours of data with a dedicated trigger
optimized to select 2-track peripheral events~\cite{janet}.  The
trigger rate of 20-40 Hz was filtered to 1-2 Hz by the final trigger,
which reconstructed the tracks on-line.  About 20,000 events were
written to tape.

\begin{figure}[t]
\centerline{\psfig{figure=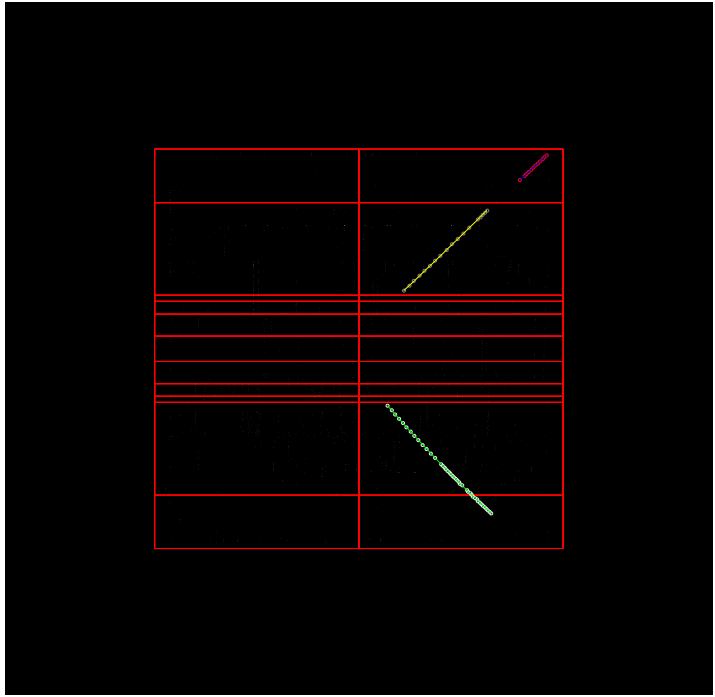,height=2.2 in,angle=270}}
\vskip .8 in
\caption{Side view of an event collected with the
peripheral collisions trigger.  The invariant mass and
$p_\perp$ are consistent with coherent $\rho^0$ production.}
\label{event}
\end{figure}

The initial event selection required a 2-oppositely-charged track,
primary vertex in the interaction diamond.  The tracks were required
to be at least slightly acoplanar to eliminate cosmic ray muons, and
the event had to have a small $p_\perp$. About 300 events passed these
cuts. This data is now being analyzed for signals from $e^+e^-$ pair
and $\rho^0$ production - the two processes with the largest cross
sections.  Figure \ref{event} shows an example of a $\rho^0$
candidate.

The CMS collaboration plans to study peripheral collisions with lead
and calcium beams at LHC\cite{cms}.  Their plans are at a
fairly early stage.

\section{Conclusions}

Peripheral collisions of heavy nuclei can probe a wide variety of
phenomena, including many faces of strong QED.  Production of $e^+e^-$
and $q\overline q$ pairs can probe the electrodynamics of the vacuum.
Besides the physics interest, peripheral collisions affect many other
areas, as a tool for hadron spectroscopy, and impacting accelerator
design,

After many years of theoretical discussion, experimental results are
beginning to become available.

\section*{Acknowledgements}

I would like to acknowledge Joakim Nystrand, my collaborator in the
studies of vector mesons.  This work was supported by the U.S. DOE,
under Contract No.  DE-Ac-03-76SF00098.

\end{document}